$\\
 \documentstyle[11pt]{article}

\newcommand{\bez}{\begin{eqnarray*}}
\newcommand{\eez}{\end{eqnarray*}}
\newcommand{\be}{\begin{equation}}
\newcommand{\ee}{\end{equation}}
\newcommand{\beq}{\begin{eqnarray}}
\newcommand{\eeq}{\end{eqnarray}}
\newcommand{\bc}{\begin{center}}
\newcommand{\ec}{\end{center}}
\newbox\grsign \setbox\grsign=\hbox{$>$} \newdimen\grdimen \grdimen=\ht\grsign
\newbox\simlessbox \newbox\simgreatbox \newbox\simpropbox
\setbox\simgreatbox=\hbox{\raise.5ex\hbox{$>$}\llap
     {\lower.5ex\hbox{$\sim$}}}\ht1=\grdimen\dp1=0pt
\setbox\simlessbox=\hbox{\raise.5ex\hbox{$<$}\llap
     {\lower.5ex\hbox{$\sim$}}}\ht2=\grdimen\dp2=0pt
\setbox\simpropbox=\hbox{\raise.5ex\hbox{$\propto$}\llap
     {\lower.5ex\hbox{$\sim$}}}\ht2=\grdimen\dp2=0pt

\title{\vspace{4cm}\large\bf
Cosmic internal symmetry in a finite universe}
\author{Arthur D.~Chernin\\
 Sternberg Astronomical Institute, Moscow University, Moscow, 119899,
Russia,\\
and Tuorla Observatory, University of Turku, Piikki\"o, FIN-21500, Finland
}

\date{~}

\begin{document}

\maketitle

\begin{abstract}
\noindent Cosmic internal symmetry (COINS) relates  vacuum (V),
dark (D) matter, baryons (B) and radiation (R). It implies
conservation of the Friedmann integral $A$ which is numerically
the same for each of the energies: $A \sim A_V \sim A_D \sim A_B
\sim A_R$. Evidence for COINS comes from the concordance data,
including the WMAP data which also indicate that the co-moving
space has a finite size. COINS 1) links the value of the vacuum
density, $\rho_V$, to the total dark mass, $M_D$, the total
baryonic mass $M_B$, and the total number, $N_R$, of
ultrarelativistic particles: $ \rho_V/M_{Pl}^4  \sim
(M_{Pl}/M_D)^{2} \sim (M_{Pl}/M_B)^{2} \sim N_R^{-4/3}$; 2) is
behind the coincidence of the observed cosmic densities at the
present epoch when the size of the universe $a(t_0) \sim A$; 3)
rules out significant deviations from spatial flatness; 4) might
fix the initial amplitude, $\Delta$, of cosmic perturbations;  5)
controls cosmic entropy $D$ (per dark matter particle); 6)
suggests a solution to the naturalness problem. The COINS nature
is due to the electroweak-scale ($M_{EW} \sim 1$ TeV) physics.
The hierarchy number, $X = M_{Pl}/M_{EW} \sim 10^{15}$, provides
the quantities involved with an appropriate universal measure: $A
\sim X^4 M_{Pl}^{-1}, \, \rho_V \sim X^8 M_{Pl}^4, \, M_D \sim
M_B \sim X^4 M_{Pl}, \, N_R \sim X^6, \, \Delta \sim X^{-2}$ (at
$M_{EW}$ temperatures),  $D \sim X$. If two extra dimensions of
the submillimeter size $R_*$ exist and eliminate the hierarchy
effect, the vacuum density $\rho_V \sim R_*^{-4}$.



\end{abstract}
\section{Introduction}

A recent analysis of the first-year data of the Wilkinson
Microwave Anisotropy Probe (WMAP) (Bennett et al. 2003, Spergel
et al. 2003) has suggested that the observed co-moving space of
the Universe is finite, containing a finite amount of matter
(Luminet et al. 2003). This is an expanding space, and its
present-day size is comparable to the current radius of the
visual horizon. The conclusion  follows from the shape of the
power spectrum of the cosmic microwave background (CMB) anisotropy
at the lowest harmonics, or largest spatial scales. It is found
that the quadrupole is only about one-seventh as strong as would
be expected in an infinite flat space. A similar effect (while
not so dramatic) is observed also for the octopole. The lack of
power on the largest scales indicates most probably that the
space is not big enough to support them. A special model -- the
Poincar\'e dodecahedral space of positive spatial curvature with
the density parameter $\Omega \simeq 1.013 > 1 $ -- is
demonstrated to reproduce the observed shape of the spectrum
better than do models of an infinite space (Luminet et al. 2003).
The figure for the density parameter  is compatible with the WMAP
limitations: $\Omega = 1.02 \pm 0.02$ (Bennett et al. 2003).

The result can be verified by further analysis of the WMAP data
and the upcoming Planck data. It may be expected that --
independently of the specific model with its rigid geometry and
exact parameter $\Omega$ -- the major conclusion about a finite
size of the real Universe will survive future tests.

Under this last assumption, I discuss below physics of a finite universe with a focus on a
special correspondence between cosmic vacuum and matter in the finite-size space. This
vacuum-matter relation is a kind of symmetry. As is well-known, symmetries are usually
classified into geometrical and internal symmetries, in fundamental physics. Geometrical
symmetries concern relations in space and time, while internal symmetries relate, for
instance, different particles of like kind. Vacuum-matter symmetry is internal symmetry that
relates vacuum and three basic forms of matter - dark matter, baryons, radiation. Like other
symmetries, this {\em cosmic internal symmetry} implies conservation of a certain quantity
which is the {\em Friedmann integral}. Friedmann integral is a time-independent genuine
characteristic of each type of cosmic energy. Its numerical value can be found with the use of
observational data. This value proves to be the same, approximately, for vacuum, dark matter,
baryons and radiation. The time-independent identity of the four values of the Friedmann
integral is a phenomenological manifestation of cosmic internal symmetry.

In Sec.2, the Friedmann integral is introduced, and its numerical estimation is made on the
basis of the recent concordance data, including the WMAP data on the CMB anisotropy power
spectrum treated a l\'a  Luminet et al. (2003); in Sec.3, a link to fundamental theory is
discussed with the use of the standard physics of the early Universe; in Sec.4, cosmic
internal symmetry is used for the study of the cosmic coincidence, the observed flatness of
space, the large-scale perturbation amplitude, the cosmic entropy; in Sec.5, this symmetry is
considered in the context of extra dimensions; summary is given in Sec.5.

\section{The Friedmann integral (FINT)}

The dynamical equation of the Friedmann cosmology is based on
differential geometry, and so it is valid independently of any
topology. It may be written for the radial size, a(t), of a
finite co-moving space in the form:
\begin{equation}
{\dot a}^2 = (A_V/a)^{-2} + A_D/a + A_B/a + (A_R/a)^{2} - k.
\end{equation}
\noindent In a finite-size universe with positive spatial
curvature (as in the model by Luminet et al. 2003), the size
$a(t)$ may be equal to the  spatial curvature radius; in this
case $k = 1$, in the Friedmann equation. Generally, the size and
the radius of curvature may be different, and in such a case,
$k=(M/M_C)^{2/3}$, where $M$ is the total nonrelativistic matter
mass (dark matter and baryons) in the universe and $M_C$ is the
matter mass within the volume with the radial size equal to the
curvature radius. For a finite-size universe of negative spatial
curvature, $k = - (M_B/M_C)^{2/3}$, in the equation above. For a
finite-size flat space, $k = 0$. (Hereafter the speed of light $c
= 1$.)

The constants $A$ in Eq.(1) represent the four major cosmic
energies which are vacuum (V), dark matter (D), baryons (B) and
radiation, or relativistic (R) energy. The constants are the
integrals of the Friedmann `thermodynamical' equation:
\begin{equation}
A = [\kappa \rho a^{3(1+w)}]^{\frac{1}{1+3w}},
\end{equation}
where $\kappa = \frac{8 \pi G}{3}$ and $G$ is the gravitational
constant; $w$ is the pressure-to-density ratio which is $-1, 0, 0,
1/3$ for vacuum, dark matter, baryons and radiation,
respectively. The integral $A$ is the {\em Friedmann integral}
(hereafter FINT).

It is obvious that the FINT defined by Eq.2 for dark matter or
baryons is directly associated with the total masses of dark
matter, $M_D$, or baryons, $M_B$, in the finite-size universe:
\begin{equation}
A_D = 2 \kappa M_D, \;\; A_B = 2 \kappa M_B.
\end{equation}
\noindent
Similarly, the FINT defined for radiation is determined by the
total number of CMB photons, $N_{CMB}$, in the finite-size universe:
\begin{equation}
A_R \propto N_{CMB}^{2/3}.
\end{equation}
\noindent As for the FINT for vacuum, it is given by the vacuum
density:
\begin{equation}
A_V = (\kappa \rho_V)^{-1/2}.
\end{equation}

Due to its origin from the Friedmann thermodynamical equation as a
constant of integration, the FINT is not restricted by any theory
constraints (except for trivial ones), and its values for various
energy forms are completely independent of each other {\it a
priori}.

The four values of the FINT can be evaluated numerically with the
use of the current concordance figures for the four energy
densities (Riess et al.1998, Perlmutter et al. 1999, Bennett
2003, Spergel 2003):
\begin {eqnarray}
\Omega_V = 0.7 \pm 0.1, \\ \Omega_D = 0.3 \pm 0.1,\\
\Omega_B = 0.02 h_{100}^{-2},  \\
\Omega_R = 0.6 h_{100}^{-2} \alpha \times 10^{-4}, \;\; 1 <
\alpha < 10-30.
\end {eqnarray}
\noindent The figures are in general agreement with the Hubble
parameter $h_{100} = 0.70 \pm 0.15$ and the cosmic age $t_0 = 14
\pm 1$ Gyr.

With these data, the FINT for vacuum
\begin{equation}
A_V = (\kappa \rho_V)^{-1/2}= \Omega_V^{-1/2} H^{-1}  \simeq 3
\times 10^{28} \; cm.
\end{equation}
The constant quantity is near the Hubble radius, $H(t_0)^{-1}$, at
the present epoch.

The three other FINT values are estimated with the use of the
 present-day size $a(t_0)$ of the finite universe. The WMAP data on the
power spectrum at lowest harmonics suggest that the size must be
near the Hubble radius:
\begin{equation}
a (t_0)  \sim H^{-1} \simeq 10^{28} \; cm.
\end{equation}

The double approximate identity $a(t_0) \simeq H^{-1} \simeq A_V $
is a specific physical characteristic of the current epoch in the
cosmic evolution.

With the size $a(t_0) \simeq H^{-1}$, one has
\begin{equation}
A_D =  \kappa \rho_D a^3 = \Omega_D a^3 H^2 \simeq \Omega_D H^{-1}
 \simeq 3 \times 10^{28} \; cm.
\end{equation}

\begin{equation}
A_B = \kappa \rho_B = \Omega_B a^3 H^2 \simeq \Omega_B H^{-1}
\simeq 3 \times 10^{27} \;cm.
\end{equation}

\begin{equation}
A_R = (\kappa \rho_R )^{-1/2} a^2 = (\Omega_R \alpha)^{-1/2} a^2 H
 \simeq 3 \times  10^{26} cm, \;\;\; (\alpha \simeq 10).
\end{equation}

Note that, instead of the value $A_R$, three values may be
introduced to the Friedman equation to account separately for the
CMB photons, $A_{CMB}$, neutrinos $A_{\nu}$, other relativistic
relics, $A_{RR}$, of the early cosmic evolution. It is obvious
that $A_{CMB} \sim A_{\nu} \sim A_R$. As for $A_{RR}$, no direct
data is yet available; one may only expect that $A_{RR} \sim A_R$,
if, for instance, relic cosmological gravitons really exist.

\section{Cosmic internal symmetry (COINS)}

As we see, all the four FINT values calculated for the finite
universe with the concordance
data prove to be identical, on the order of magnitude:
\begin{equation}
A_V \sim A_D \sim A_B \sim A_R \sim A \sim 10^{60 \pm 1} M_{Pl}^{-1}.
\end{equation}
Here units are used in which the speed of light, the Boltzmann
constant and the Planck constant are all equal to unity: $c = k =
\hbar = 1$. The Planck mass is $M_{Pl} = G^{-1/2} \simeq 1.2
\times 10^{19}$ GeV.

The time-independent identity of Eq.15  may be treated as a
symmetry relation that brings vacuum into correspondence with
matter (as it was earlier proposed, --  Chernin 2001, 2002a).
Vacuum-matter symmetry does not relate to space and time, and
therefore it is a type   of internal symmetry. Symmetry between
the proton and the neutron in nuclear physics is a basic example
of internal symmetry: the particles are different in mass,
electric charge, etc., but characterized by the same charge in
nuclear (strong) interaction. The {\em cosmic internal symmetry}
(hereafter COINS) is the symmetry of the cosmic energies: they
are different in physical state, content, temporal behavior,
etc., but each of them is characterized by the same
(approximately) quantity which is the FINT.

COINS comes from the phenomenological analysis of the available observational data,
and it reveals most naturally, as is seen from Sec.2, in the finite-size universe.

COINS is not an exact symmetry; it is valid on the order of
magnitude and violated at a level of a few percent, on the
logarithmic scale: $ \lg (A_V/A_R)/\lg A_V \simeq 0.03$, with $A$
in the Planck units.(According to Okun (1985), the notion of
symmetry is essentially close to the notion of beauty.
Furthermore, true beauty in its supreme forms requires a slight
deviation from symmetry which communicates to beauty a mysterious
and alluring element of non-finito.) COINS itself and its
violation are both important for cosmological problems discussed
below (Sec.4).

COINS is a covariant symmetry: it is formulated in terms of the FINT which is a  scalar
(invariant) quantity. FINT is associated with the four-dimensional Riemann invariant, $R = 8
\pi G (\rho - 3 p)$. In the limit of infinite time $R \rightarrow 32 \pi G \rho_V = 12/A_V^2,
\; t \rightarrow \infty$. If cosmic matter is initially generated in the form of massless
particles (and the particles acquire mass later via, say, the Higgs mechanism), the invariant
is the same in the opposite time limit as well: $R \rightarrow 12/A_V^2, \; t \rightarrow 0$.

COINS is a time-independent symmetry in the evolving Universe; it exists unchanged during all
the time when the observed cosmic energies exist themselves. Like most of symmetries, COINS
corresponds to the conservation of the appropriate quantity: this is obviously the FINT which
is a genuine `charge' of this symmetry.

COINS is a robust symmetry: it is not sensitive to (not too big) variations of the figures
involved. For instance, the FINT for the non-vacuum energies may be re-estimated with the use
of another, than in Sec.2, value of $a(t_0)$. It may be equal, say, to the curvature radius in
the model with the present-day $\Omega = 1.013$,  as suggested by Luminet et al.(2003):
\begin{equation}
a (t_0) = a_C (t_0) = (\Omega -1)^{-1/2} H^{-1} \simeq 10 H^{-1}
\simeq 10^{29} \;cm.
\end{equation}
\noindent
Then one will have, instead of Eq.15:
\begin{equation}
A_V \sim A_D \sim A_B \sim A_R \sim A \sim 10^{62 \pm 1} M_{Pl}^{-1}.
\end{equation}
An approximate identity of the four values of the FINT takes place for the new figures as well.

The physical nature of COINS can be clarified and the symmetry
relation for vacuum, dark matter and radiation can be explained
by (or even deduced from) a freeze-out model (Chernin 2001,
2002a). The model is based on the standard physics of the early
universe and assumes that the weakly interacted dark matter
particles (WIMPs) have a mass near the the fundamental electroweak
energy scale $m_D \sim M_{EW} \sim 1$ TeV. Freeze-out of dark
matter annihilation occurs at temperatures $\sim m_D \sim 1$ TeV
when the red shift $z \sim M_{Pl}/M_{EW}$. The model provides
COINS with a link to fundamental physics: it indicates that the
FINT can be represented in terms of the two fundamental energies,
which are $M_{Pl}$ and $M_{EW}$, only:
\begin{equation}
A  \sim (\bar M_{Pl}/M_{EW})^4 M_{Pl}^{-1} \simeq 10^{61 \pm 1} M_{Pl}^{-1}.
\end{equation}
Here the reduced Planck scale $\bar M_{Pl} = g M_{Pl}, \, g \simeq 0.1-0.3$, is introduced
to account (as usual) for some dimensionless factors (like $8 \pi/3$, etc.) of the theory.

According to the model, the constant vacuum density,
\begin{equation}
\rho_V \sim (\bar M_{Pl}/M_{EW})^{-8} M_{Pl}^{4} \simeq 10^{-122 \pm 2} M_{Pl}^4,
\end{equation}
is expressed in terms of the same fundamental energies (see also Arkani-Hamed et al. 2000).

The ratio $\bar M_{Pl}/M_{EW} \sim 10^{15}$ that enters the result is known as
the {\em hierarchy number} in particle physics. It characterizes the huge
gap between the two fundamental energies; the nature of the gap is not well understood,
and this is considered as one of the central problems  in fundamental theory.
We address this problem in Sec.4 below.

Note that the FINT for baryons is not treated by the freeze-out model. It may, however, be
assumed that the physics of baryogenesis is basically behind the relation between the FINT for
baryons and the FINT for dark matter (Chernin 2001). Baryogenesis and associated physics might
also be responsible (at least, in part) for the COINS violation.

\section{Solving problems with COINS}

Let us turn now to some old and new problems in cosmology and show how cosmic
internal symmetry may help in their better understanding.

\subsection{Cosmic coincidence}

Why are the observed densities of vacuum, dark matter, baryons
and radiation are nearly coincident? This is the {\em cosmic
coincidence problem} which has appeared with the discovery of
cosmic vacuum (Riess et al.1998, Perlmutter et al. 1999) and is
usually considered as a severe challenge to current cosmological
concepts (see, for instance, Chernin 2002a and references
therein). As is well-known, the  idea of quintessence was
introduced in an attempt to eliminate the problem. However it is
now clear that quintessence can hardly be useful because the
pressure-to-density ratio has recently been found to lie between
-1.2 and -0.9 (Perlmutter et al. 2003), which seemingly rules out
the idea.

On the contrary, the approach based on COINS offers a natural solution to the problem without
any additional assumptions. Indeed, one may easily see that the four densities can become
identical (approximately), because the four values of the FINT are identical (approximately):
\begin{equation}
\kappa \rho_V \sim A^{-2}, \; \kappa \rho_D \sim (A/a)^3, \; \kappa \rho_B \sim (A/a)^3, \;
\kappa \rho_R \sim (A/a)^4.
\end{equation}
The densities happen to coincide now-days, because $a(t_0) \sim A$ right now.

Thus, the cosmic densities are observed to be nearly coincident, because of COINS and the
special character of the moment of observation at which $a(t_0) \sim A$.

\subsection{The Dicke problem}

The geometry of the co-moving space looks nearly flat in observations. Why this is so? The {\em
problem of flatness} was recognized by Dicke (1970) and formulated in terms of the
time-dependent parameter $\Omega (t)$. As Dicke (1970) mentioned, the universe must be
extremely finely tuned to yield the observed balance between the total energy density of the
Universe and the critical density. This balance must be tuned with the accuracy $\sim
10^{-16}$ or $\sim 10^{-60}$, if it is fixed at the epoch of the light element production or
at the Planck epoch, respectively.

COINS shows the problem in quite different light. With the use of simple relations based on
the Friedmann equation of Sec.2 (see also Chernin 2003) one may find that the maximal possible
deviation from $\Omega = 1$ in a universe with the space of positive or negative curvature
occurs at $z = a(t_0)/({\frac{1}{2}}A_V^2 A_D)^{1/3}-1 \simeq 1$. The deviation goes to zero
in both limits $t \rightarrow 0$ and $t \rightarrow \infty$, as is seen directly from the
Friedmann equation.

The maximal possible deviation is expressed in term of the FINT values for vacuum and dark
matter:
\begin{equation}
\vert \Omega_{ex} -1 \vert \simeq \vert
[1 - \frac{1}{2}( \frac{A_V}{A_D})^{2/3}]^{-1} -1 \vert,
\end{equation}
in the case when the finite space has positive curvature and its size is the curvature radius
(see Sec.2). The most severe WMAP limits, $\Omega -1 = 0.02 \pm 0.02$, are met, if $A_V/A_D
\le 0.008$. This is possible when $a(t_0) \ge (5-10) H^{-1}$. The model proposed by Luminet et
al. (2003) is a concrete example which fits this condition.

In a more general case, one has, in accordance with Sec.2,
\begin{equation}
\vert \Omega_{ex} -1 \vert \simeq \vert
[1 \pm \frac{1}{2} ( \frac{A_V}{A_D})^{2/3} (M/M_C)^{2/3}]^{-1} -1 \vert,
\end{equation}
\noindent It is seen from here that a finite space of positive curvature fits the WMAP data,
if, say, $A_V/A_D \simeq 1$ and the present-day curvature radius $a_C (t_0)$ is 5 times the
radial size of the space $a (t_0)$.

To feel the contrast with the fine-tuning argument, one may
compare the modest numbers between 3 and 10 with the enormous
numbers
 $10^{-16}$ or $10^{-60}$.

Thus, the interplay between vacuum antigravity and dark matter gravity at $z \sim 1$ is behind
the observed near flatness of the space. The interplay is controlled by COINS,  and  COINS
rules out completely any significant deviations from flatness  at present, in the past and
future of the Universe. The exact quantitative measure of a possible non-flatness is
determined by both COINS itself and its violation, -- without any fine tuning.

Note that no hypothesis (about, say, an enormous vacuum density,
or enormous energy density of inflanton field, at very large $z$)
is required to understand the physics of the Dicke problem. The
really observed vacuum density and simple cosmology at modest $z$
are quite enough to understand why the observed space looks nearly
flat.

\subsection{Perturbation amplitude}

Cosmic perturbations that give rise to structure formation are characterized by their spectrum
and initial amplitude. The spectrum can be obtained in the theory of quantum fluctuations
developed first by Chibisov and Mukhanov (1981); they obtain the spectrum which is similar in
shape to the Harrison-Zeldovich spectrum and well confirmed by the WMAP and other recent data.
However, there is no theory yet that could give the perturbation initial amplitude.

In the theory of the perturbation evolution, one faces another fine-tuning problem which
nevertheless seems to be fairly similar to that in the Dicke argument. Indeed, the
perturbations must be extremely finely tuned in amplitude to come to the nonlinear regime
between the red shifts, say, $z \simeq 3-10$ and $z \simeq 1$ (at $z < 1$ vacuum antigravity
terminates the perturbation growth).

Consider, for instance, an example (Chernin 2002b) of large-scale
adiabatic perturbations which are ever over-horizon in size and
ever grow before $z \sim 1$. Such scales are larger than 10-30
Mpc at the present epoch (and may also be near the largest scales
related to the low-harmonic end of the CMB anisotropy power
spectrum -- see Sec.1). If the perturbations are generated  at
the epoch of light element production or at the Planck epoch,
their initial amplitudes (estimated with the standard theory)
must be tuned with the accuracy  $10^{-16}$ or $10^{-60}$,
respectively.

COINS suggests a new approach to the problem. To start with, one
may consider the large-scale perturbation as above and use, in
addition, an idea by Zeldovich (1965). In his study of weak
adiabatic perturbations, Zeldovich found that the correct time
rate of the perturbation growth could be obtained in a simple
picture in which a perturbation was treated as an area of
negative total energy on the unperturbed background of the
parabolic expansion, -- if to use the language  of the Newtonian
mechanics. In other words, any such perturbation may be
considered as a part of a universe with a co-moving space of
positive curvature. Assume also that the FINT values are the same
in the perturbation area and in the background model. Then,
extending the Zeldovich idea in this way, one may find that the
perturbations come indeed to the nonlinear regime just at about
$z \sim 1$.

To see this, one may address again the relations of the subsection above. Indeed, the relative
amplitude of density perturbations is obviously related to the density parameter $\Omega$, in
the Zeldovich picture:
\begin{equation}
\delta \rho/\rho \simeq \Omega -1,
\end{equation}
It reaches its maximum value at $z \sim 1$, and at that time
\begin{equation}
\delta \rho_D/\rho_D = \Omega_{ex} -1 \simeq [1 - \frac{1}{2}( \frac{A_V}{A_D})^{2/3}
(M/M_C)^{2/3}]^{-1} -1.
\end{equation}
The value is about unity, $\delta \rho_D/\rho_D \sim 1$, provided
$(A_V/A_D) (M/M_C) \sim 1$. The latter requirement may, for
instance, has the form: $A_V/A_D \sim 1, \, M/M_C \sim 1$.

As we see, the perturbations become nonlinear at $z \sim 1$ simply because of COINS. No fine
tuning in amplitude is needed, and no (explicit) assumption about the perturbation amplitude
is involved in the considerations. The basic assumption is only the validity of COINS in both
perturbation and background. If perturbations are generated in this special way, their correct
time behavior is guaranteed: they reach the nonlinear regime on time.

(It must be, however, assumed also that the perturbation area and the background start to
expand at more or less the same moment of time. It would be enough, actually, if the
difference in the start time, $\delta t$, is, say, several times less than the cosmic age at
the epoch of $z \sim 1$. The effect of asynchronous start diminishes with time as $\delta
t/t$; this is the so-called falling perturbation mode, which is not too significant in any
reasonable circumstances.)

The perturbation evolution and its time-dependent amplitude are
given by Eq.23; one may use it to estimate the amplitude of this
type perturbations at, say, the freeze-out temperature $\sim
M_{EW}$  (see Sec.3) when the red shift $z = z_{EW} \sim \bar
M_{Pl}/M_{EW}$:
\begin{equation}
\Delta (z_{EW}) \simeq \Omega (z_{EW}) -1 \simeq [a(t_0)/A_R]^2(1 + z)^{-2}
\sim (M_{EW}/\bar M_{Pl})^2 \sim 10^{-30}.
\end{equation}
\noindent

Thus, COINS shows how initial perturbations might be prepared
without any fine tuning. The freeze-out model associated with
COINS gives the perturbations a natural initial amplitude in
terms of the two fundamental energies: this is the inverse
squared the hierarchy number, at the electroweak temperatures.

Remind that the considerations concern  the perturbations of ever over-horizon spatial scales.
But in their physical nature, the perturbations on these scales are hardly different
essentially from the perturbations on all the other scales. The theory by Chibisov and
Mukhanov (1981) gives rather an  evidence for a common nature of the perturbations over their
whole scale range. If so, physics which is responsible for the origin of the cosmic
perturbations involves COINS and the hierarchy effect as its basic ingredients.

\subsection{Cosmic entropy}

The number density of the CMB photons $n_R \sim 1000$, at
present. The baryon number density $n_B \sim 10^{-6}$ now. The
time-independent ratio, $B = n_R/n_B \sim 10^9$, is called the Big
Baryonic Number; as is well-known, it  represents the cosmic
entropy per baryon. Why this number is so big? This question is
referred to as the {\em cosmic entropy problem}.

In terms of the FINT, the Big Baryonic Number may be represented as
\begin{equation}
B \sim A_R^{3/2} A_D^{-1} m M_{Pl}^{-1/2}.
\end{equation}
If to use COINS as an exact symmetry and put $A_R = A_B$, one has from here $B \sim 10^{11}$,
which is not too bad as the first approximation. If not only symmetry, but its violation as
well is taken into account, the number will obviously be quite correct.

The freeze-out model suggests that the {\em Big Dark Number} may also be of interest:
\begin{equation}
D = n_R/n_D \sim 10^{12},
\end{equation}
where $n_D$ is the number density of dark matter particles and it
is assumed again that the WIMP mass $m_D \sim M_{EW}$. In terms of
the FINT, one has
\begin{equation}
D \sim A_R^{3/2} A_D^{-1} M_{EW} M_{Pl}^{-1/2}.
\end{equation}
For exact symmetry relation, $A_R = A_D$, and with the FINT
expressed via the two fundamental energies, this gives
\begin{equation}
D \sim \bar M_{Pl}/M_{EW} \sim 10^{15},
\end{equation}
which is the simplest (and perhaps `more fundamental' than $B$) measure for the cosmic entropy
per particle.

Thus, in the first approximation, one may answer the question above:  the cosmic entropy per
particle is given by a big number because of COINS and the hierarchy effect in fundamental
physics. COINS and the hierarchy control -- via cosmic entropy -- cosmic light element
production in the Big Bang.

\section{COINS, hierarchy and extra dimensions}

As we see, the hierarchy number provides COINS and the associated
phenomena with  a common quantitative measure: the basic
constants of cosmology which are the FINT, cosmic entropy per
particle, the initial perturbation amplitude are expressed via
this big number. Implicitly this number is also involved in the
phenomena of cosmic coincidence and flatness. Therefore, if the
hierarchy problem is resolved in fundamental physics, it will
give a new insight into the nature of COINS.

The hierarchy problem has been discussed in fundamental physics
for decades. We will not consider here earlier ideas and address
rather a recent one by Arkani-Hamed et al. (1998). This is the
idea of macroscopic extra dimensions which are proposed to
eliminate the energy hierarchy of fundamental theory.

The idea assumes that there exist finite (compactified)
submillimeter  extra dimensions in space, and all the dimensions
constitute a close multi-dimensional space. This
multi-dimensional space is treated as the {\em true space of
nature}. It is also assumed that there is one and only one {\em
truly fundamental} energy scale $M_*$ in nature, and  it is close
to the electroweak scale $M_{EW}$. As for the Planck scale, it is
reduced to a combination of the scale $M_*$ and the size $R_*$ of
the compact macroscopic extra dimensions of the true space:
\begin{equation}
M_{Pl} \sim (M_{*} R_*)^{n/2} M_{*}.
\end{equation}
\noindent Here $n$ is the number of the spatial extra dimensions, which are proposed to be of
the same size. It is reasonably argued  that the case $n = 2$ is the most appropriate one; if
so, the size of two extra dimensions is in the millimeter (or rather submillimeter) range:
\begin{equation}
R_* \sim 0.1 \;\;cm, \;\;\; n = 2.
\end{equation}

Together with the Planck mass, the gravitational constant in three-dimensional space, $G =
M_{Pl}^{-2}$, looses its fundamentalism and is reduced to the two truly fundamental constants
$M_*$ and $R_*$.

We will not go here into further details of the extra-dimension
physics and adopt from it only one result: the hierarchy number,
$ M_{Pl}/M_{EW}$, is replaced with the product $ M_* R_*$. This
is actually not elimination of the hierarchy, but its
re-formulation in the new terms of $M_*$ and $R_*$.

In the multi-dimensional space, all the physical fields, except gravity, are assumed to be confined in the three-dimensional space, or brane. Therefore the Friedmann theory (together with the FINT, COINS, $B$, $D$, $\Delta$, etc.) is defined at the {\em cosmological brane} of finite sizes. The multi-dimensional physics affects the brane, and therefore cosmology  should be re-formulated in terms of the true fundamental constants. In this way, one has for the FINT:
\begin{equation}
A \sim (M_{*} R_*)^{(3/2)n} M_{*}^{-1}.
\end{equation}
\noindent
In the case of two extra dimensions:
\begin{equation}
A \sim (M_{*} R_*)^{2} R_*, \;\;\;\; n = 2.
\end{equation}

Then the vacuum density
\begin{equation}
\rho_V  \sim (M_{*} R_* )^{-2n} M_{*}^{4},
\end{equation}
and in the case of two extra dimensions:
\begin{equation}
\rho_V \sim R_*^{-4}; \;\; n = 2.
\end{equation}

This is the most surprising result: the vacuum density proves to be expressed via the size of
the extra dimensions alone. The new relation is free from any signs of the hierarchy effect
(Chernin 2002c). This is a case when the hierarchy is really eliminated from the
multi-dimensional physics.

According to the idea of  extra dimensions, all we observe in
three-dimensional space are shadows of the true multi-dimensional
entities. In particular, it may be assumed that true vacuum
exists in the multi-dimensional space, and the observed cosmic
vacuum is not more than its 3D projection to the cosmological
brane. This is possible, if vacuum is due to gravity only and not
related to the fields of matter. In this case, the true vacuum is
defined in the multi-dimensional space, and its density $\rho_{V5}
\sim R_*^{-6}$, in the case of two extra dimensions. This true
vacuum is also free from the hierarchy effect. Via vacuum, COINS
brings the energies on the brane in a correspondence with the
physics of the multi-dimensional world.

Turning to the topics of Sec.3 above, one finds for the Big Dark Number:
\begin{equation}
D \sim M_* R_*, \;\;\; n = 2.
\end{equation}

In a similar way, the initial (at the electroweak temperatures)
amplitude of the large-scale perturbations may be found:
\begin{equation}
\Delta \sim (M_* R_*)^2, \;\;\; n = 2.
\end{equation}

Taking the relations at face, one may conclude that the basic
cosmological parameters have roots in the extra-dimension physics,
-- if extra dimensions really exist. It is expected that their
existence will be directly tested with the TeV hadron collider and
in submillimeter laboratory experiments in several years --
perhaps, at the same time when the Planck mission will test
compactness on cosmological scales.

\section{Discussion and Conclusions}

Cosmic internal symmetry (COINS) is a time-independent phenomenon
in the evolving Universe. COINS unifies vacuum and matter,
revealing their internal correspondence and perhaps common
physical nature. Evidence for the existence of COINS comes in a
natural way from the concordance observation data. The evidence
is most straightforward, if the size of the co-moving space
Universe is finite. The WMAP data in the interpretation by
Luminet et al. (2003) do suggest that the Universe is finite,
containing a finite matter mass.

COINS imply that the vacuum density $\rho_V$ is in a direct quantitative inter-relation with
the total mass of dark matter, $M_D$, the total mass of baryons, $M_B$, and the total number
of ultrarelativistic particles (first of all, the CMB photons and neutrinos), $N_R$, in the
finite Universe:
\begin{equation}
\rho_V M_{Pl}^{-4}  \sim (M_{Pl}/M_D)^{2} \sim (M_{Pl}/M_B)^{2} \sim N_R^{-4/3}.
\end{equation}

The freeze-out model based on the standard physics provides COINS with a link to fundamental
theory. As a result, the dimensionless hierarchical number, $X = \bar M_{Pl}/M_{EW} \sim
10^{15}$, emerges and gives all the quantities involved a universal measure:
\begin{equation}
A \sim X^4 M_{Pl}^{-1}, \;\; \rho_V  \sim X^{-8} M_{Pl}^4, \;\;
M_D \sim X^4 M_{Pl}, \;\; M_B \sim X^{4}M_{Pl}, \;\; N_R \sim X^{6}.
\end{equation}

If the hierarchy effect of fundamental physics roots in (two) macroscopic extra dimensions
(Arkani-Hamed et al. 1998), the true energy $M_* \sim 1$ TeV and the submillimeter size of
extra dimensions $R_*$ come to the cosmology scene; then $X \sim M_* R_*$, and the relations
take the form:
\begin{equation}
A \sim X^{2} R_*, \;\; \rho_V  \sim R_*^{-4}, \;\;
M_D \sim X^5 M_{*}, \;\; M_B \sim X^{5}M_{*}, \;\;N_R \sim X^{6}.
\end{equation}

It has been long recognized (see Okun, 1985, and references therein) that the energy scale $M_*
\sim 1$ TeV plays a central part in fundamental physics. Now COINS shows that, in this way or
another, the energy proves to be most significant as well in cosmology.

Other results of the discussion  may be briefly summarized as
follows:

* The present-day size of the expanding space is equal or close, on the
order of magnitude, to the value of the FINT: $a(t_0) \sim A$. This is a
special feature of the present epoch of the cosmic evolution.

* COINS is behind the coincidence of the four observed cosmic density, and the phenomenon takes
place just now, because $a(t_0) \sim A$ at present.

* COINS controls the interplay between matter gravity and vacuum antigravity in space-time;
as a result, it rules out any considerable deviations from spatial flatness now, in the past,
and in the future of the Universe.

* COINS might be involved in the generation mechanism for cosmic perturbations; it implies
that the perturbation initial (at the epoch of $\sim 1$ TeV temperatures) amplitude $\Delta
\sim X^{-2}$, is given in terms of the hierarchy number.

* COINS is responsible for cosmic entropy, so that entropy per dark matter particle, the Big
Dark Number, $D = n_R/n_D \sim X$, is also measured by the
hierarchy number. Indirectly, via entropy, COINS controls the
outcome of the cosmic light element production in the Big Bang.

* Finally, COINS suggests a solution to the long standing {\em naturalness problem}. It shows
that the vacuum density is as it is, because all the values of
the FINT,  including the value for vacuum, are identical. In
addition, COINS together with extra-dimension physics lead to a
fairly  natural (and the simplest) relation: $\rho_V \sim
R_*^{-4}$.

In principle, almost all the results above may equally be
formulated for any reasonable cosmological model compatible with
the current concordance data in both finite or infinite space.
Specific data may be used as well. For instance, the WMAP figure
for $\Omega = 1.02 \pm 0.02$ suggests a close space, and the
corresponding present-day curvature radius ($\simeq 6 \times
10^{28}$ cm) may easily replace the size $a(t_0)$ in the
relations of Secs.2-3. This will not, however, alter the symmetry
relation or change further results in any significant way.
Another version is the simplest flat-space model with infinite
space; in this case, the FINT evaluation can be made with the use
the co-moving size of the Metagalaxy, the currently visible part
of space, ($\simeq 10^{28} (1+z)^{-1}$ cm ), -- and again it will
not change the results. In this sense, COINS is a robust and
model-independent result of cosmic phenomenology.

\section*{References}

Arkani-Hamed N. et al., 1998, PhL B429, 263

Arkani-Hamed N. et al., 2000, PhRvL 85, 4434

Bennet C.L. et al., 2003, ApJ Suppl. Ser. 148, 1

Chernin A.D., 2001, Phys.-Uspekhi 44, 1099

Chernin A.D., 2002a, New Asdtron. 7, 113

Chernin A.D., 2003b, astro-ph//0211489

Chernin A.D., 2003c, astro-ph//026179

Chernin A.D., 2003, New Astron., 8, 79

Chibisov G. \& Mukhanov V., 1981, JETP Lett. 33, 532

Dicke R., 1970, {\it Gravitation and the Universe}, Amer. Phil. Soc., Philadelphia, PA

Luminet J.-P. et al., 2003, Nature 425, 593

Okun L.B., 1985, {\it Particle Physics}. Harwood, N.Y.

Perlmutter S. et al., 1999, ApJ 517, 565

Perlmutter S. et al., 2003, astro-ph//03099368

Ries A. et al., 1998, AJ 116,

Spergel D.N. et al., 2003, ApJ Suppl. Ser. 148, 175

Zeldovich Ya.B., 1965, Adv. Astron. Astrophys. 3, 241

\end{document}